\documentclass[preprint,12pt]{elsarticle}

\usepackage{graphicx}
\usepackage{subfigure}
\usepackage{caption}
\usepackage{epstopdf}
\usepackage{epsfig}
\usepackage{amssymb}
\usepackage{amsthm}
\usepackage{ulem}

\journal{Solid State Communications}

\begin{document}
\begin{frontmatter}


\title{ 3$d$ - 4$d$ hybridization anomaly in Ni$_x$Pd$_{1-x}$ alloys at quantum critical point}


\author{P. Swain, Sanjeev K. Srivastava}
\address{Department of Physics, Indian Institute of Technology
Kharagpur, Kharagpur-721302, INDIA}

\author{Suneel K. Srivastava}
\address{Department of Chemistry, Indian Institute of Technology
Kharagpur, Kharagpur-721302, INDIA}

\begin{abstract}
{\it First-principles} density functional theory computations of electronic structure and local magnetic properties of the non-fluctuating ground state of Ni$_x$Pd$_{1-x}$ alloy system around its quantum critical point $x_c = 0.026$ have been performed. The density of states at the Fermi energy and certain other parameters characterizing the Ni 3$d$ - Pd 4$d$ hybridization apparently follow power-laws with $x$ similar to that obeyed by the reported ferromagnetic to paramagnetic transition temperature. The width of Pd $4d$ density of states (DOS) and centroid of Ni 3$d$ DOS show peak-like anomalies in the neighbourhood of $x_c$, and so indicate a possible scenario of the existence of a definite relation between the orbital hybridization and the emergence of quantum fluctuations in the system.

\end{abstract}


\end{frontmatter}

\section{Introduction}
\label{S:1}

Quantum phase transitions (QPT's) \cite{Zhong, Lohneysen, Zhang, Nicklas99}, i.e. transitions taking place at zero temperature, have recently become a subject of intense research in theoretical and experimental condensed matter physics. These transitions are driven by zero temperature quantum fluctuations and are accessed by varying a non-thermal control parameter $q$, like chemical composition, magnetic field, pressure, etc. \cite{Sachdev-aX}, at a critical value $q_c$, known as quantum critical point (QCP). Physical properties associated with quantum fluctuations are quite distinct from those associated with thermal fluctuations responsible for conventional temperature-driven phase transitions. For example, in metallic systems, the material at QCP shows an unconventional non-Fermi liquid (NFL) behaviour \cite{Nicklas99}. The existence of such critical points has been proved to hold the key to many technologically exploitable phenomena, e.g. the occurrence of high-temperature superconductivity \cite{Sachdev_Sci}, behaviour of two-dimensional electron gases \cite{Kravchenko}, etc. Perhaps the most easily understandable class of a QPT is the ferromagnet-to-paramagnet (FM-PM) transition at 0 K in itinerant electron systems as a function of the strength of exchange interaction between electron spins \cite{Belitz}. The FM-PM QPT occurring in metallic Ni$_x$Pd$_{1-x}$ alloy system across its QCP ($x_c$ = 0.026) is one such example, wherein signatures of quantum criticality have been observed in bulk macroscopically \cite{Nicklas99} and microscopically \cite{Srivastava}, and in nanodimensions macroscopically \cite{swain15}.

Several theories, including quantum Ising model, quantum rotor model, Hubbard model and Heisenberg-spin model, with renormalization group approaches and space-time scalings, exist for understanding QPT's and associated critical behaviours of different kinds of systems undergoing QPT \cite{Sachdev_bk, MVojta}. Computational studies of QPT's, however, are posed with various formidable challenges including (i) the problem of quantum many-particle, (ii) occurrence of the transition only in the thermodynamic limit of infinite system size, and (iii) anisotropic space-time scaling at QCP \cite{TVojta}. The first of these, i.e. the quantum many-particle problem, has the possibility to be addressed with sophisticated approximation methods like Hartree-Fock or density functional theory (DFT) \cite{Hohenberg, Kohn, Mishra, Mohanta}. These theories, however, neglect fluctuations, which are crucial for understanding phase transitions including QPT, and hence would fail in describing the quantum fluctuations associated with QPT's \cite{TVojta}. Nonetheless, it is still intriguing to explore what DFT computations of the non-fluctuating ground state say vis-a-vis the QPT occurring in a material system. DFT results have a relevance also in the case of the usage of Hubbard model, where these are to be used as an initial input. Although some computational methods, like quantum Monte Carlo, density matrix renormalization group and dynamical mean field theory have been proposed for finding QCP and predicting quantum critical behaviour \cite{TVojta}, none have hitherto been utilized for computations to the best of authors' knowledge. Thus, using DFT for computations of ground state properties of systems undergoing a QPT and comparing the results with experiments would perhaps be the first, although still very small, step in this direction.

In this scenario and with the above motivation, we take in this work the Ni$_x$Pd$_{1-x}$ alloy system to represent the class of materials exhibiting QPT, compute the ground state electronic and local magnetic properties of the system across the QCP by DFT, and seek any correlation with the experimental quantum-fluctuation driven properties reported in the literature \cite{Nicklas99}.

\section {Calculational details}

The full-potential linearized augmented plane wave (FLAPW) method of DFT as implemented in the code Wien2K \cite{Blaha, Cottenier} was employed to perform the present calculations. The generalized gradient approximation as introduced by Perdew, Burke and Ernzerhof \cite{Perdew} was taken as the exchange-correlation functional. Pure Pd and Ni crystal structures were constructed in the cubic Fm$\bar{3}$m space group. The lattice parameters were then optimized by calculating the total energy at different volumes and fitting the result to the Birch Murnaghan equation of state \cite{Birch, Murnaghan} for both Pd and Ni. In order to construct the structures of Ni$_x$Pd$_{1-x}$ alloys with $x \neq 0, 1$ and in the vicinity of $x_c$, the following strategy has been adopted: out of very many possible combinations $n, l, m$ of generating Pd supercells of sizes $n \times l \times m$, only the smallest supercell volume combinations have been taken in such a manner that replacing just the (0, 0, 0) Pd by Ni suffices to create the particular Ni$_x$Pd$_{1-x}$ alloy. The purpose of taking the smallest supercells has been to minimize the computational cost. Since the study is a comparative one, computational accuracy for any individual $x$ is not very crucial, and hence this strategy of computation should be good enough. The $x$ values and the corresponding supercells are listed in Tab. 1. The equilibrium lattice constants for each of the supercells were once again obtained by the same method as for Pd and Ni. Further, the atomic coordinates were relaxed in all the impurity cases to reduce the atomic forces to less than 1 mRy/au. In the FLAPW method, the basis function is set differently in two regions -  non overlapping muffin-tin sphere of radius R$_{\rm MT}$, and the remaining interstitial region - around an atom. In the muffin-tin region, the potential is taken to be atom-like and atomic spherical wave functions are used. In the interstitial region, on the other hand, the electrons behave like plane waves. In present calculations, the R$_{\rm MT}$ value for both Ni and Pd atoms are taken to be 2.5 au. The maximum multipolarity of the spherical wavefunctions has been taken as 10, while the length $\rm {K_{max}}$ of the maximum wavevector for the plane-wave basis set has been limited by $\rm {R_{MT} \times K_{max}}$ = 7.0. The charge density was Fourier expanded up to $\rm {G_{max}}$ = 12. For sampling of the Brillouin zone, a k-mesh with 110 irreducible k-points in the irreducible wedge of the Brillouin zone was used. For each concentration both unpolarized and spin-polarized self-consistent field calculations were performed.

\begin{table}
\caption{\label{tab1}Values of Ni concentration $x$ and the
corresponding supercell size ($n \times l \times m$). } \centering
\begin{tabular}{lll}
\hline
$x$  &supercell size & \\
\hline
0.02    &$4 \times 3 \times 1$ and $2 \times 2 \times 3$& \\
0.027   &$3 \times 3 \times 1$& \\
0.03    &$4 \times 2 \times 1$ and $2 \times 2 \times 2$&\\
0.04    &$3 \times 2 \times 1$& \\
0.06    &$2 \times 2 \times 1$& \\
 \hline
\end{tabular}\\

\end{table}

\section {Results and Discussion}

The first step of analysis is to see which of the non-magnetic and ferromagnetic states is more stable for a particular $x$ according to the calculations, and whether the inferences are consistent with the existing experimental reports \cite{Nicklas99}. For this, the difference $\Delta \rm{E}$ between the total energies from the spin-polarized (E$_{\rm {sp}}$) and unpolarized (E$_{\rm {un}}$) calculations ($\Delta \rm{E} = {\rm {E_{sp}}} - {\rm {E_{un}}}$) for each composition has been calculated and plotted against $x$ in Fig. 1. Two features can be identified in the figure: (i) All the points corresponding to $n \times l \times 1$ supercells vary more or less smoothly with $x$, while if $m \neq 1$ cases are categorized separately, the point corresponding to the $2 \times 2 \times 2$ supercell of this category is scattered far away. For this reason, it would not be inappropriate to consider just the $n \times l \times 1$ supercells for further analyses. Therefore, only the $n \times l \times 1$ supercells are being analysed hereafter. (ii) For small $x$, the points corresponding to the $n \times l \times 1$ supercells can be shown to follow a $x^{3/4}$ dependence, as displayed in the inset of Fig. 1. This is almost the same as the variation $T_c \sim (x - 0.026)^{3/4}$ of the FM-PM transition temperature $T_c$ with $x$ as reported by Nicklas {\it et al.} \cite{Nicklas99}. With these kinds of highly simplified structures, approximating $(x - 0.026)^{3/4}$ to $x^{3/4}$ must be acceptable. This gives further strength to the strategy of taking only $n \times l \times 1$ supercells for further analyses.

\begin{figure}%
\centering
\includegraphics[width=0.6\textwidth]{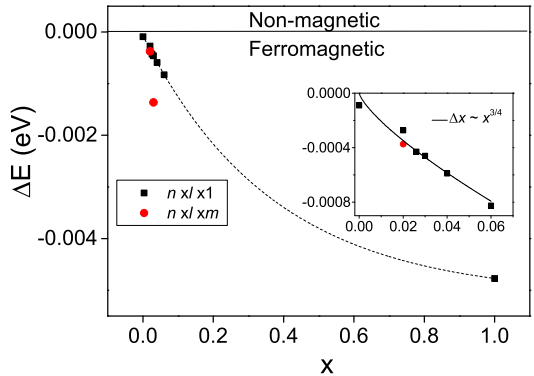}%
\hspace{8pt}%

\caption[]{Variation of the energy difference $\Delta \rm{E}$ with $x$ for various supercells. Inset: Small-$x$ portion of the curve along with a $x^{3/4}$ fit.}%
\end{figure}

Figure 2a shows the total densities of states (TDOS's) from unpolarized calculations for all the studied compositions with $n \times l \times 1$ supercells (Tab. 1). An examination of the figure suggests that on incremental introduction of Ni, the shape of the Pd ($x = 0$) TDOS gets modified systematically in such a way that each of the three major peaks, as well as the whole TDOS, broadens. This is an indication of some hybridization between certain Pd and Ni orbitals \cite{Rakesh}. This aspect will be investigated further later. At this point, it would be interesting to see the behaviour of the `TDOS at the Fermi energy' $N({\rm E_F}$) with $x$, as displayed in Fig. 2b. From the figure, a power-law dependence of N(E$_{\rm F}$) on $x$ similar to that of $T_c$ on $x$ as reported by Nicklas {\it {et. al.}} \cite{Nicklas99} can be observed beyond $x = 0.02$. Assuming the Stoner exchange integral $I$ for the onset of ferromagnetism \cite{ Vosko} to be slightly less than the 1/$N{\rm(E_F}$) value for pure Pd, i.e., $I \sim 1/2.62 \sim 0.38$, we can say that ferromagnetism appears in the system at $x = 0.02$ and gets stronger according to a power law with $x$, in agreement once again with the existing report of dependence of $T_c$ on $x$ \cite{Nicklas99}. The assumed value $\sim 0.38$ of $I$ is not different from the value (0.39) reported by Hong and Lee \cite{Hong}. This way, taking $n \times l \times 1$ supercells for the analyses is further justified.

\begin{figure}%
\centering
\subfigure[][]{%
\includegraphics[width=0.35\textwidth]{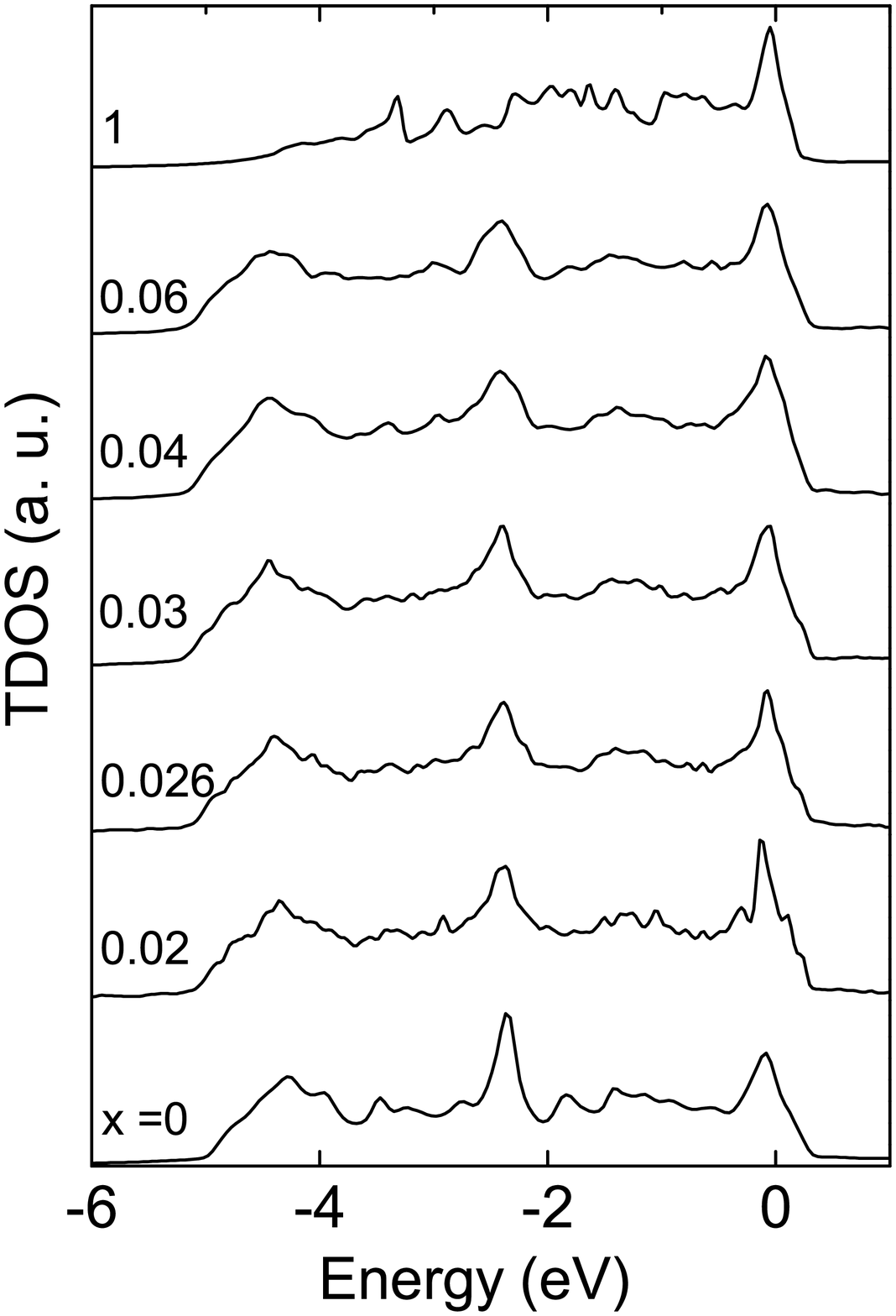}}%
\hspace{8pt}%
\subfigure[][]{%
\includegraphics[width=0.45\textwidth]{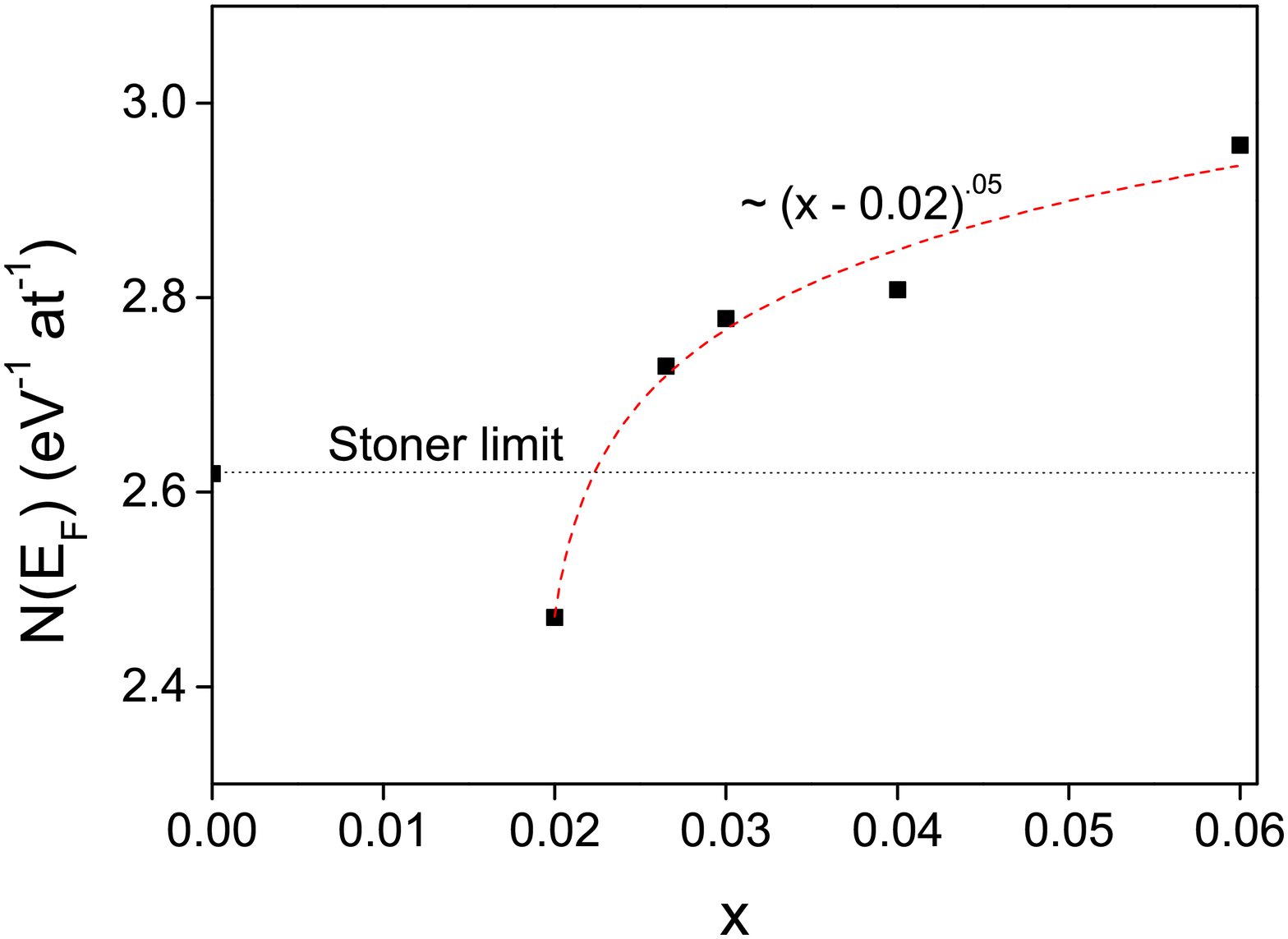}}%
\hspace{8pt}%

\caption[]{(a) Total densities of states from unpolarized calculations for all the studied compositions with $n \times l \times 1$ supercells. (b) Variation of N(E$_{\rm F}$) with $x$ along with the power-law fit. The estimated Stoner limit is indicated. }%
\end{figure}

The TDOS's from spin-polarized calculations are shown in Fig. 3a. Qualitatively the same inferences can once again be drawn for each spin component as for the corresponding unpolarized cases. It would further be interesting to see the $x$ dependences of the magnetic moments on Ni and its nearest neighbour Pd, as derived from the spin-polarized calculations and shown in Fig. 3b. An overall increasing trend of the moments on both the atoms with incremental addition of Ni to Pd can be observed for small Ni concentrations from the figure. This observation can be understood as follows: Ni as an impurity in Pd host is known to induce ferromagnetic spin-polarization of Pd conduction electrons and give rise to giant moments in its vicinity \cite{Cheung, Beille}. This way, moments on both Pd and Ni are likely to increase with Ni doping on an average for low Ni concentrations. However, a smooth variation may not be obtainable from the calculations (as is observable from Fig. 3b) because the model structures taken in the calculations are far different from and far simpler than real experimental samples \cite{Beille}.

\begin{figure}%
\centering
\subfigure[][]{%
\includegraphics[width=0.35\textwidth]{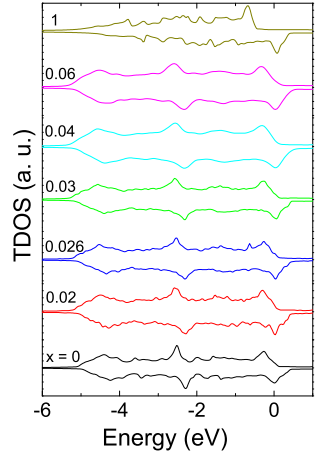}}%
\hspace{8pt}%

\centering
\subfigure[][]{%
\includegraphics[width=0.45\textwidth]{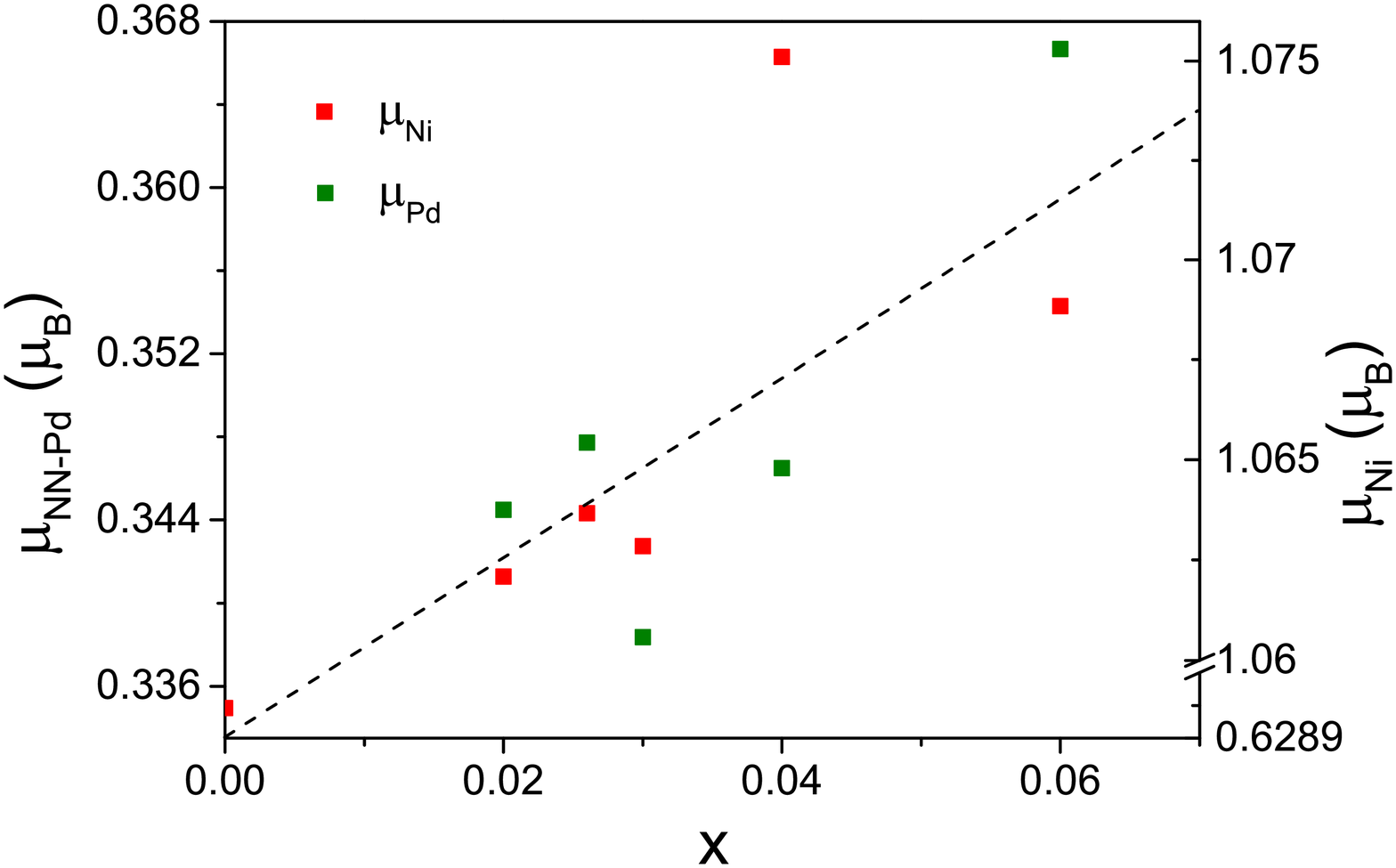}}%
\hspace{8pt}%
\caption[]{(a) Total densities of states from spin-polarized calculations for all the studied compositions with $n \times l \times 1$ supercells. (b) Variation of impurity Ni and NN Pd moments with $x$. The dotted straight line indicates the increasing trend of the moments with $x$.}%
\end{figure}

We would now investigate the afore mentioned hybridization between certain Pd and Ni orbitals. According to the calculations, the local DOS's of Ni and Pd are dominated by 3$d$ and 4$d$ partial DOS's (PDOS's), respectively (not shown). It would thus be sufficient to compare the 3$d$ PDOS of impurity Ni and 4$d$ PDOS of its nearest neighbour (NN) Pd. The Pd 4$d$ PDOS's for $x \neq 1$ and Ni 3$d$ PDOS's for $x \neq 0$ are shown in Fig. 4a. That there are variations in both the types of PDOS's with $x$, which in turn are indicative of Ni 3$d$-Pd 4$d$ hybridization, is apparent at the first sight of the figure. For further analysis, we adopt the method of semi-quantifying hybridizations from DOS's as used in our previous work \cite{Rakesh}. For this, we determine the width ($w_{\rm Ni}$), centroid ($c_{\rm Ni}$) and height ($h_{\rm Ni}$) of the Ni 3$d$ PDOS's and the width ($w_{\rm Pd}$) and centroid ($c_{\rm Pd}$) of the NN Pd 4$d$ PDOS's, and then look at their variations with $x$. An increase in hybridization then would be signalled as a concurrent widening and lower energy shifts of the centroids of the two PDOS's, and simultaneous increase and decrease of the width and height of a particular PDOS \cite{Rakesh}. Plots of $w_{\rm Ni}$ and $w_{\rm Pd}$ versus $x$, $w_{\rm Ni}$ and $h_{\rm Ni}$ versus $x$, and $c_{\rm Ni}$ and $c_{\rm Pd}$ versus $x$ are shown in Figs. 5a, 5b and 5c, respectively. The above mentioned variations, characteristics of an increasing Ni 3$d$-Pd 4$d$  hybridization with $x$, can be readily observed from the figures. This leads to the first outcome of the study that the Ni 3$d$-Pd 4$d$  hybridization increases monotonically on introducing Ni impurities incrementally into Pd matrix. Existence of Ni 3$d$-Pd 4$d$  hybridization in diluted $Pd$Ni alloys has been reported earlier in a photoemission study \cite{Acker}, while such $x$-dependence has not been studied.

\begin{figure}%
\centering
\includegraphics[width=0.40\textwidth]{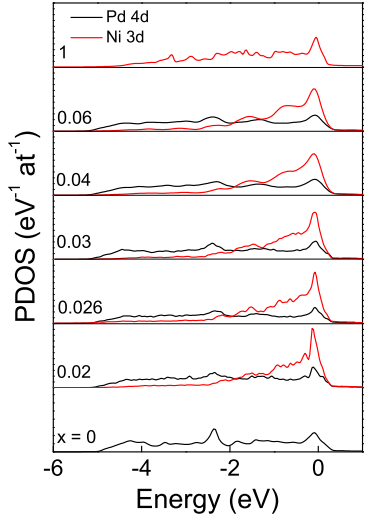}%
\hspace{8pt}%

\caption[]{Impurity Ni 3$d$ and NN Pd 4$d$ partial densities of states.}
\end{figure}

However, there is another interesting observation associated with all the curves in Figs. 5a - c: apart from the systematic variations with $x$ as mentioned above, each curve possesses an anomaly in the vicinity of $x_c = 0.026$. A close observation suggests that $w_{\rm Ni}$ has a somewhat steep rise around $x_c$ and then increases with a power-law like trend, if a linear background is subtracted. This can be likened with the power-law $x$ dependences of $T_c$ and $N{\rm(E_F)}$ and hence may have a correlation with the occurrence of magnetic order in Pd on Ni doping. A concurrent but opposite trend is shown by $h_{\rm Ni}$ and $c_{\rm Pd}$. This reverse nature of the two quantities is quite likely from the hybridization point of view as discussed above. However, $w_{\rm Pd}$ and $c_{\rm Ni}$ exhibit a rather sharp peak and a similar dip, respectively, around the QCP, over and above the quite apparent monotonic power-law-like increase or decrease with $x$. These peak-like anomalies certainly can not arise due to the presence of quantum fluctuations at the QCP, since such fluctuations are not implementable in the calculations. But, then there is no explanation for the occurrence of these exceptions. Keeping in mind the smooth variation of the total energy of the supercells and the concurrent behaviours of other quantities with the onset of magnetic order, the anomalies can perhaps also not be discarded as mere artefacts. This way, there appears to be a definite relation between the orbital hybridization of the non-fluctuating ground state and the emergence of quantum fluctuations in the Ni$_x$Pd$_{1-x}$ alloy system. Exactly what this relation could be can however not be determined or explained at this stage.

\begin{figure}%
\centering
\includegraphics[width=0.40\textwidth]{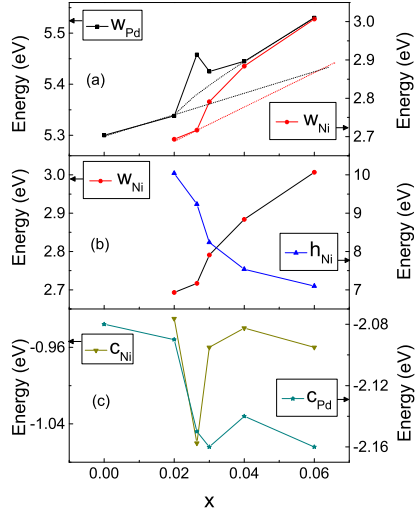}%
\hspace{8pt}%

\caption[]{Variations of (a) widths $w_{\rm Pd}$ and $w_{\rm Ni}$, (b) width $w_{\rm Ni}$ and height $h_{\rm Ni}$, and (c) $c_{\rm Pd}$ and $c_{\rm Ni}$, with $x$. Power-law behaviours and the $w_{\rm Pd}$ peak in (a) are schematically indicated; for others, these features are qualitatively apparent and hence are not shown.}%
\end{figure}

\section {Conclusion}

We have computed electronic and local magnetic properties of the non-fluctuating ground state of Ni$_x$Pd$_{1-x}$ alloy system across the quantum critical point by {\it first-principles} density functional theory. The simplest model structures constructed with the smallest possible $n \times l \times 1$ Pd supercells, the (0, 0, 0) atom of which is replaced by Ni, have been chosen for the study. The primary reason to choose the $n \times l \times 1$ structural constructions has been that the total energy for these varies smoothly with $x$. The density of states at the Fermi energy follows a power-law with $x$ similar to that obeyed by the experimental ferromagnetic to paramagnetic transition temperature. The local magnetic moments on the impurity Ni and nearest neighbour Pd atoms have a monotonic increase with $x$. The study gives rise to two main outcomes (i) the Ni 3$d$-Pd 4$d$  hybridization increases monotonically, and according to the occurrence of magnetic order, on introducing Ni impurities incrementally into Pd matrix, and (ii) the hybridization is anomalous at the QCP, which is indicative of a possible scenario of the existence of a definite relation between the orbital hybridization and the emergence of quantum fluctuations in the Ni$_x$Pd$_{1-x}$ alloy system.


\begin{thebibliography}{99}

\bibitem{Zhong} W. Zhong and D. Vanderbilt, Phys. Rev. B {\bf 53} (1996) 5047.

\bibitem{Lohneysen} H. v. L\"ohneysen, Physica B {\bf 206 - 207} (1995) 101.

\bibitem{Zhang} X. Zhang, C-L. Hung, S-K. Tung, N. Gemelke and C. Chin, New J. Phys. {\bf 13} (2011) 045011.

\bibitem{Nicklas99} M. Nicklas, M. Brando, G. Knebel, F. Mayr, W. Trinkl and  A. Loidl, Phys. Rev. Lett. {\bf 82} (1999) 4268.

\bibitem{Sachdev-aX} S. Sachdev and B. Keimer, {\it arxiv.org/pdf/1102.4628v2.pdf}.

\bibitem{Sachdev_Sci} S. Sachdev, Science {\bf 288} (2000) 468.

\bibitem{Kravchenko} S. V. Kravchenko, W. E. Mason, G. E. Bowker, J. E. Furneaux, V. M. Pudalov and M. D'lorio, Phys. Rev. B {\bf 51} (1995) 7038.

\bibitem{Belitz} D. Belitz and T. R. Kirkpatrick, J. Phys.: Cond. Matter {\bf 8} (1996) 9707.

\bibitem{Srivastava} S. K. Srivastava and S. N. Mishra, Physics Teacher {\bf 50} (2009) 10.

\bibitem{swain15} P. Swain, Suneel. K. Srivastava and Sanjeev. K. Srivastava, Phys. Rev. B {\bf 91} (2015) 045401.

\bibitem{Sachdev_bk} S. Sachdev, Quantum Phase Transitions 2nd Ed. (Cambridge: Cambridge University Press) 2011.

\bibitem{MVojta} M. Vojta, {\it arxiv.org/pdf/cond-mat/0309604.pdf}.

\bibitem{TVojta} T. Vojta, {\it arxiv.org/ftp/arxiv/papers/0709/0709.0964.pdf}.

\bibitem{Hohenberg} P. Hohenberg and  W. Kohn,  Phys. Rev.  {\bf 136} (1964) B864.

\bibitem{Kohn} W. Kohn and L. J. Sham, Phys. Rev. {\bf 140} (1965) A1133.

\bibitem{Mishra} S. N. Mishra and S. K. Srivastava,  J. Phys.: Cond. Matter {\bf 20} (2008) 285204.

\bibitem{Mohanta} S. K. Mohanta, S. N. Mishra and S. K. Srivastava, J. Magn. Magn. Mat. {\bf 355} (2014) 142.

\bibitem{Blaha}P. Blaha, K. Schwarz, G. Madsen,  D. Kvasnicka and  J. Luitz 1999 Wien2K: an Augmented Plane Wave + Local Orbitals Programme for Calculating Crystal Properties, Techn. Universit\"at Wien, Austria, ISBN 3-9501031-1-2.

\bibitem{Cottenier}S. Cottenier 2002 Density Functional Theory and the Family of (L)APW-methods: A Step-By-Step Introduction (K.U.Leuven, Belgium: Instituut voor Kern-en Stralingsfysica), ISBN 90-807215-14.

\bibitem{Perdew} J. P. Perdew,  K. Burke and M. Ernzerhof,  Phys. Rev. Lett. {\bf 77} (1996) 3865.

\bibitem{Birch} F. Birch,  Phys. Rev. {\bf 71 } (1947) 809.

\bibitem{Murnaghan} F. D. Murnaghan,  Finite Deformation of an Elastic Solid (New York: Wiley) 1951.

\bibitem{Rakesh} R. Das, G. P. Das, and S. K. Srivastava, J. Phys. D: Appl. Phys. {\bf 49 } (2016) 165004.

\bibitem{Vosko} S. H. Vosko and J. P. Perdew,  Can. J. Phys. {\bf 53} (1975) 1385.

\bibitem{Hong} S. C. Hong and J. I. Lee, J. Korean. Phys. Soc. {\bf 52} (2008) 1099.

\bibitem{Cheung} T. D. Cheung, J. S. Kouvel, and J. W. Garland, Phys. Rev. B {\bf 23} (1981) 1245.

\bibitem{Beille} J. Beille and G. Chouteau, J. Phys. F: Metal Phys. {\bf 5} (1975) 721.

\bibitem{Acker} J. F. van Acker {\it et al.}, Phys. Rev. B {\bf 38} (1988) 10463.




\end{thebibliography}
\end{document}